\documentclass[aps,pre,twocolumn,showpacs,showkeys,a4paper,nofootinbib]{revtex4}

\usepackage{amsmath} 
\usepackage{amsfonts} 
\usepackage{amsthm}
\usepackage{bm}
\usepackage{graphicx}
\usepackage{epsfig}
\usepackage[all,2cell]{xy}
\usepackage{color}

\usepackage{stmaryrd}

\usepackage{amscd}
\usepackage{amssymb}
\usepackage{fullpage}
\usepackage{color}

\usepackage{float}
\usepackage{empheq} 
\usepackage{fancybox}
\usepackage{pdfpages}

\usepackage{epsf}
\usepackage{epsfig}
\usepackage{epstopdf}

\newcommand{\be}{\begin{equation}}
\newcommand{\ee}{\end{equation}}
\newcommand{\bea}{\begin{eqnarray}}
\newcommand{\eea}{\end{eqnarray}}
\newcommand{\bes}{\begin{subequations}\bea}
\newcommand{\ees}{\eea\end{subequations}}
\newcommand{\ba}{\begin{array}}
\newcommand{\ea}{\end{array}}

\newcommand{\hhat}[1]{#1}
\newcommand{\var}{\mathrm{var}\,}
\newcommand{\mean}{\mathrm{mean}\,}

\newcommand{\cur}{J}

\begin{document}

\newtheorem{theorem}{Theorem}


\title{Tightening the uncertainty principle for stochastic currents} 

\author{Matteo Polettini} 
 \email{matteo.polettini@uni.lu}
\affiliation{Complex Systems and Statistical Mechanics, University of Luxembourg, Campus
Limpertsberg, 162a avenue de la Fa\"iencerie, L-1511 Luxembourg (G. D. Luxembourg)} 

\author{Alexandre Lazarescu}
\affiliation{Complex Systems and Statistical Mechanics, University of Luxembourg, Campus
Limpertsberg, 162a avenue de la Fa\"iencerie, L-1511 Luxembourg (G. D. Luxembourg)}

\author{Massimiliano Esposito}
\affiliation{Complex Systems and Statistical Mechanics, University of Luxembourg, Campus
Limpertsberg, 162a avenue de la Fa\"iencerie, L-1511 Luxembourg (G. D. Luxembourg)}
 
\date{\today}

\begin{abstract}
We connect two recent advances in the stochastic analysis of nonequilibrium systems: the (loose) uncertainty principle for the currents, 
which states that statistical errors are bounded by thermodynamic dissipation; and the analysis of thermodynamic consistency of the currents in the light of symmetries. Employing the large deviation techniques presented in [Gingrich et al., Phys. Rev. Lett. 2016] and [Pietzonka et al., Phys. Rev. E 2016], we provide a short proof of the loose uncertainty principle, and prove a tighter uncertainty relation for a class of thermodynamically consistent currents $ \cur$. Our bound involves  a measure of partial entropy production, that we interpret as the least amount of entropy that a system sustaining current $ \cur$ can possibly produce, at a given steady state. We provide a complete mathematical discussion of quadratic bounds which allows to determine which are optimal, and finally we argue that the relationship for the Fano factor of the entropy production rate $\var \sigma / \mean\, \sigma \geq 2$ is the most significant realization of the loose bound. We base our analysis both on the formalism of diffusions, and of Markov jump processes in the light of Schnakenberg's cycle analysis.
\end{abstract} 


\maketitle

\section{Introduction}

After Heisenberg's uncertainty principle in Quantum Mechanics was formulated, a large variety of uncertainty relations have also been derived in statistical mechanics, based on statistical concepts such as the Fisher information and the Shannon entropy \cite{gilmore,luo,birula}. Today, a mature theory of thermodynamics is available, based on the solid mathematics of stochastic processes and the physical principles of Stochastic Thermodynamics \cite{broeck,seifert1,jarzynski}. Several authors have then inspected the statistical properties of the fundamental observables of Stochastic Thermodynamics, namely the currents, allowing them to first substantiate \cite{barato1,barato2} and then prove \cite{barato3,gingrich} a general nonequilibrium uncertainty principle, which roughly states that, in a nonequilibrium process, ``the least the error, the most the dissipation''. A refinement of this statement using a non-quadratic bound was also conjectured in Ref.\,\cite{barato3} and later proven in Ref.\,\cite{pietzonka2}. Moreover, a similar inequality holds between the dissipation, and the average time of an estimation of the arrow of time \cite{roldan}.

One remarkable feature of Stochastic Thermodynamics is that it puts propositions from statistical physics in a physical perspective, in this case the theory of large deviations \cite{touchette} of random variables defined along long-time realizations of a Markovian process.  {Long-time observables are of two kinds: some measure static properties of the process (e.g. the typical number of cars peering at a crossroad); others measure dynamical properties such as currents (e.g. the net number of cars through a street). As regards Markov processes,  static observables in the long-time limit depend only on the steady density $\rho_\star$. This is the crucial object at equilibrium, where there are no currents nor dissipation.  Nonequilibrium thermodynamics, instead, is involved both with static observables and, most importantly for this paper, with the behaviour of some current $\cur$. In this context, the stochastic uncertainty relation states that
\bea
\frac{\var  \cur}{(\mean  \cur) ^2} \; \geq \frac{2}{\sigma_\star}, \label{eq:eprbound}
\eea 
where $\sigma_\star$ measures the steady-state dissipation rate (in units of the Boltzmann's constant per time; from here on $k_B=1$), and $\sqrt{\var  \cur}/\mean  \cur$ is the error. This inequality was first proposed by Barato and Seifert in Ref.\,\cite{barato1} in the context of Markov jump processes, and therein derived for cycle-currents of a network in the linear regime (slightly out of equilibrium), and for unicyclic networks arbitrarily far from equilibrium. Large deviation inequalities  {based on the steady density $\rho_\star$} were then provided in Refs.\,\cite{barato3,gingrich}, allowing Pietzonka et al. to conjecture useful bounds \cite{barato3}, and Gingrich et al. \cite{gingrich} to provide a full, and quite involved, proof for generic currents; similar inequalities have also been derived for a case of a driven periodic diffusion \cite{nyawo}. As a first contribution we provide in Sec.\,\ref{quadratic} a simpler and more general proof, valid for all stochastic processes that verify a certain mathematical property (of which jump processes and diffusions are examples), and which highlights the crucial role played by the Gallavotti-Cohen symmetry.

Not all current-like observables are amenable to physical interpretation, and furthermore, as we will argue, the bound expressed in Eq.\,(\ref{eq:eprbound}) comes from a quadratic approximation of the rate function that is not optimal. In Ref.\,\cite{polettini}, some of the authors  {of the present paper} proposed a theory of {\it thermodynamic consistency} of the currents. For a current-like observable to be consistent, a corresponding symmetry of the thermodynamic driving forces must be obeyed. This prompts us to inquire the question whether a tighter bound holds for thermodynamically consistent currents.

In this paper we generalize the treatment to overdamped diffusion processes (Sec.\,\ref{diffusions}). In particular, we analyze a class of thermodynamically consistent currents $ \cur^a$, for which we can prove the tighter bound
\bea
\frac{\var  \cur^a}{(\mean  \cur^a) ^2} \; \geq \frac{2}{\sigma_\star^{a}} \label{eq:tightbound}
\eea  
 {where $\sigma^{a}_\star$ is the minimum entropy production rate that can be achieved by a system that sustains current $ \cur^a$, compatibly with a given steady density $\rho_\star$}. The class of currents for which this result holds are defined in such a way that the steady-state constraint  $\nabla \jmath = 0$ is satisfied, which involves the microscopic state-space currents in terms of which all current-like observables can be expressed as linear functionals. The analysis naturally leads to the identification of a ``nonequilibrium response matrix'', which in the linear regime allows us to connect directly to the results of Ref.\,\cite{barato1}.

Finally, we extend the analysis to Markov jump processes, to connect to previous literature and call into play Schnakenberg's network theory of macroscopic observables. With the aid of an example, we then argue that the entropy production rate itself, which is one special case of a current-type observable, is optimal with respect to the loose bound.

\paragraph*{Notations:} The asterisk $\star$ is reserved to steady-state quantities. We assume Einstein's convention on index contraction. Indices are lowered with the Kronecker symbol $\delta_{ij}$, the Euclidean scalar product is denoted $\langle\,\cdot\,\rangle$. In the case of diffusions, the divergence operator is $\nabla = \partial_i = \partial/\partial x_i$. We omit explicit dependencies whenever unnecessary. The scalar product of two vector fields is
\bea
\langle v,w \rangle = \int dx \, v_i(x)  w_j(x) \, \delta^{ij}. 
\eea  

\section{Tightening the quadratic bound on large deviations of currents}
\label{quadratic}

In this section, we provide a general understanding of the mathematical origin of the stochastic uncertainty relations, as well as a simple general proof of their validity. The results apply in particular to diffusions and to Markov jump processes.

Any observable macroscopic current $J$ is a linear combination of microscopic currents $\jmath$ whose steady-state statistics is described by a large deviation rate function $I(\jmath)$ with a minimum at $\jmath_\star$, which we assume to satisfy a Gallavotti-Cohen symmetry $I(\jmath)-I(-\jmath)=- \langle \jmath,f \rangle$, where $f$ are the conjugate forces, such that $\langle \jmath_\star,f\rangle=\sigma_\star$. The stationary density $\rho_\star$ will be fixed throughout. The full information given by that symmetry is that the antisymmetric part of $I(\jmath)$ is linear with a slope $-\frac{1}{2}f$. We can therefore decompose $I(\jmath)$ into a linear part and a symmetric part $F(\jmath)$.

Obtaining a proper quadratic bound on $I(j)$ is equivalent to finding a positive symmetric matrix $A$ (i.e. a metric) and a constant $b$ such that
\begin{equation}
F(\jmath)\leq \langle\jmath, A \jmath\rangle+b
\end{equation}
with the conditions
\begin{align}
&A\jmath_\star=\frac{1}{4}f\\
&b=\frac{1}{4}\sigma_\star
\end{align}
so that the bound is minimal and vanishes at $\jmath_\star$. This gives us $I(\jmath)\leq \langle(\jmath-\jmath_\star), A(\jmath-\jmath_\star)\rangle$, as expected. With no other requirement for $A$ than this, we have immediately that 
\begin{equation}
I(\alpha \jmath_\star)\leq \frac{\sigma_\star}{4}(1-\alpha)^2,
\end{equation}
which is all we need to prove the loose bound: that relation implies the same inequality between the second derivatives of the functions around $\alpha=1$, and by the usual arguments exposed in the mentioned references, from this equation one can obtain the loose bound Eq.\,(\ref{eq:eprbound}) for any macroscopic current, once one recognizes $I''(J) = (\mathrm{var}\,J)^{-1}$ (see  below the specific cases of diffusions and jump processes for full detail).

A sufficient condition for $A$ to exist is that $\alpha\frac{{\rm d}^3}{{\rm d}\alpha^3}F(\alpha \jmath_\star)\leq 0$, which ensures that $F$ is smaller than any osculating even parabola in the direction of $\jmath_\star$. This turns out to be the case for jump processes, from the fact that it is true for a Poisson process and that the property is stable under linear combination. It is also trivially the case for a diffusion, in which case $F$ is purely quadratic. Note that unlike the quantum uncertainty relations, which are inherent to how conjugate pairs of variables are defined in quantum mechanics, the stochastic ones are not always true: they would not hold, for instance, for a noisy Fokker-Planck equation with conserved quartic noise, however unphysical that would be.

The least precise solution for $A$ can then be constructed as an orthogonal matrix with eigenvalue $\sigma_\star$ in the direction of $\jmath_\star$ and infinity in all other directions:
\begin{equation}\label{Isigma}
  I_\star(\jmath)\leq\begin{cases}
    \frac{\sigma_\star}{4} \left(1-\alpha \right)^2, & \text{if $\exists~\alpha\in\mathbb{R}$, $\jmath=\alpha \jmath_\star$}.\\
    +\infty, & \text{otherwise}.
  \end{cases}
\end{equation}
This is the solution conjectured by Pietzonka and al. \cite{barato3} in the equivalent form of an inequality on scaled cumulant generating functions: all the bounds given are functions of a single scalar $z\jmath_\star$, where $z$ is the quantity conjugate to the current $j$ through a Legendre transform. The bound being a function of a scalar variable, it is invariant under shifts of $z$ which are orthogonal to $\jmath_\star$, and that invariance is translated into a constraint $\jmath\propto \jmath_\star$ for the large deviation function.

This observation leads to a few remarks. First of all, it is not surprising that only the total entropy production $\sigma_\star$ enters the loose bound (\ref{Isigma}), since, for a fluctuation of the form $\jmath=\alpha \jmath_\star$ for the current, all the microscopic entropy productions (edge-wise or cycle-wise) fluctuate by the same factor $\alpha$ and cannot be differentiated. Moreover, this bound can always be found because $\jmath=\alpha \jmath_\star$ is always divergence-free. However, it will be a bad bound for most contracted currents: the less our kernel $\phi$, as defined in Eq.\,(\ref{eq:phia}), projects onto $j_\star$, the less precise the bound is, and in particular, currents which are balanced on average (components of the current which vanish on average) are completely uncertain in that respect.

In the case of diffusions, it is easy to find a better quadratic bound: the large deviation function of the currents is already quadratic itself, so no approximation is needed.

For jump processes, a better solution has been found by Gingrich and al. \cite{gingrich} by choosing $A$ diagonal in the basis of edge currents, and proving the inequality for that choice (which is solved by taking $A_{ee}=\frac{f^e}{4\jmath_\star^e}$).

However, $A$ does not need to be diagonal with respect to edge currents, and in most cases, one can construct a quadratic bound strictly better than that one by considering $\langle\jmath, A \jmath\rangle-F(\jmath)$ in the space orthogonal to $\jmath_\star$, and minimise it with respect to the component of $\jmath$ along $\jmath_\star$. That function is positive, vanishes at $0$ but usually nowhere else, and increases fast enough to be bounded from below by a bilinear form. We show in Appendix \ref{appendix2} that the problem of finding an optimal bound reduces to that of finding hyperellipses inscribed in a convex manifold with one fixed contact point. We can always find at least one solution, which will typically have $2d$ contact points if $d$ is the dimension of our cycle currents space. The physical meaning of that optimal bound and of the contact points is unclear.

\section{Diffusions \label{diffusions}}

We consider a diffusion process in continuous state space, described by the following overdamped stochastic differential equation, interpreted in It\=o's calculus\footnote{We follow here the treatment of Maes and co-workers, as didactically exposed in Ref.\,\cite{wynants}. The drift correction term $\partial_j g^{ij}$ is conventionally added to avoid its appearance in later expressions (in particular in the Fokker-Planck equation). However, as detailed in Ref.\,\cite{polettiniJSTAT}, it would be desirable to add another term $g^{ij} \partial_j \ln \sqrt{\det g}$ which would grant the general covariance of the theory under coordinate transformations. However, since this term contributes a gradient to the thermodynamic force, its thermodynamic contribution is a boundary term that can be safely omitted in the forthcoming discussion.}
\bea
dx^i_t = \left[ \mu^i(x_t) + \partial_j g^{ij}(x_t) \right] dt + \sqrt{2} \, e^i_n(x_t) dw^n_t,
\eea
with nondegenerate diffusion tensor given by 
\bea
g^{ij} := e^i_{m} e^j_{n} \delta^{mn},
\eea
where $ \mu^i$ is the driving field and $e^i_n$ is the amplitude of the Gaussian noise with increment $dw^n_t$.

If we could trace an infinite number of particles evolving by the above equation, we could describe them by the probability of finding a particle in a neighbourhood of $x$ at time $t$, whose density $\rho_t(x)$ evolves by the Fokker-Planck (FP) equation
\bea
\partial_t \rho_t + \nabla \jmath_{\rho_t} = 0
\eea 
with the FP-current defined in terms of the probability density as
\bea
\jmath_{\rho} := \mu \rho - g \nabla \rho .
\eea

We focus on steady states. We assume that the FP equation is ergodic, with a unique steady density $\rho_\star$. Then the steady current $\jmath_\star := \jmath_{\rho_\star}$ is divergenceless
\bea
\nabla \jmath_\star = 0.
\eea
We further define the conjugate thermodynamic force \cite{sei}
\bea
f_i & := & \frac{g^{-1}_{ij} \jmath_\star^j}{\rho_\star} \label{eq:forcedef} \\
 & = & g^{-1}_{ij} \mu^j - \partial_i \ln \rho_\star.
\eea
At a steady state, the system delivers entropy to the environment at rate
\bea
\sigma_\star = \langle \jmath_\star,f \rangle.\label{eq:EPR}
\eea

\subsection{Macroscopic currents}
\label{macros}

Macroscopic currents are defined as linear functionals of the (microscopic) FP-currents
\bea
 \cur^a = \langle \jmath,\phi^a \rangle \label{eq:phia}
\eea
where $\phi^a_i(x)$ are some kernels, which play the crucial role of bridging the microscopic description to the macroscopic one. We assume for simplicity that these functionals are linearly independent (otherwise, macroscopic conservation laws would ensue). 

The system is {\it thermodynamically consistent} if there exist macroscopic thermodynamic forces $F_a$ such that
\bea
f_i(x) = F_a \phi^a_i(x), \label{eq:sym}
\eea
after which the entropy production rate can be written just in terms of the macroscopic quantities as
\bea
\sigma =  F_a  \cur^a.
\eea
In analogy to the treatment of discrete-state systems proposed in Ref.\,\cite{polettini}, we call Eq.\,(\ref{eq:sym}) a {\it symmetry} of the thermodynamic forces. We provide an example of a system that has thermodynamically consistent currents and forces in Appendix \ref{OU}. 

 At a steady state, given that $\jmath_\star$ is divergenceless, by Eq.\,(\ref{eq:forcedef}) then necessarily
\bea
F_a \, \partial_j  \left( \rho_\star g^{ji}  \phi_i^a \right)= 0. \label{eq:sss}
\eea
Another immediate consequence is the following relationship between the steady-state currents and the macroscopic thermodynamic forces
\bea
 \cur^a_\star =  G^{ab} F_b \label{eq:linear}
\eea
where
\bea
G^{ab} = \int  \rho_\star  g^{ij}  \phi^a_i \phi^b_j. 
\eea
Eq.\,(\ref{eq:linear}) has the form of a linear-response relationship between macroscopic currents and forces, of the kind that ensues close to equilibrium. Then, $G$ would play a role akin to the response matrix, an analogy that will be useful to interpret the upcoming results. However, it must be emphasized that the above equation is {\it not} a linear-response relationship, because $G$ itself is sensitive to small pertubations, for example, of the drift $\mu$. This nondissipative contribution to nonequilibrium response has been analyzed in terms of the activity \cite{maesactivity,maesresponse}. In the following we call $G$ the {\it nonequilibrium response matrix}.

For later use, let us introduce the quadratic {\it dissipation function}}
\bea
\sigma[\cur^1,\ldots,\cur^n] = G^{-1}_{ab} \cur^a \cur^b, \label{eq:dissifun}
\eea
 {such that $\sigma[\cur_\star] = \sigma_\star$. Variations of $\sigma[\cur_\ast]$ with respect to the current correspond to variations of the entropy production rate {\it at fixed response matrix}, which can be achieved by fixing the steady-state distribution and the diffusion matrix.

\subsection{Large deviations}

We will now derive the uncertainty principle for thermodynamically consistent currents in the context of overdamped diffusion processes. We closely retrace the discussion of Ref.\,\cite{gingrich}, from which we abundantly borrow. Our analysis allows to appreciate certain subtleties concerning thermodynamic consistency and the steady state constraint $\nabla \jmath =0$. 

Above, our discussion regarded ideal quantities such as the density traced by an infinite number of realizations of a stochastic process. Here we will consider one single realization of such a stochastic process, in a large-enough time window $[0,T$]. We are interested in certain stochastic observables, in particular the stochastic density $\rho_T(x)$ (also known as empirical measure), counting the average number of times a trajectory passes by $x$, and the empirical current $\jmath^i_T(x)$ denoting in which direction a stochastic trajectory proceeds as it passes by $x$. They are formally defined as
\bea
\rho_T(x) & = & \frac{1}{T} \int_0^T \delta(\hhat{x}_t - x) dt \\
\jmath^i_T(x) & = & \frac{1}{T} \int_0^T \delta(\hhat{x}_t - x)\circ d\hhat{x}^i_t,
\eea
where $\circ$ denotes the Stratonovich differential.
In particular, we are interested in the statistics of one particular macroscopic current marked ``1''
\bea
 \cur^1_T = \langle  \jmath_T,\phi^1 \rangle.  \label{eq:macro}
\eea
For the moment we assume that the current is not {\it orthogonal} to the steady currents, that is, $ \cur^1_\star = \langle \phi^{1},  \jmath_\star \rangle  \neq 0$.

In many situations, including the present one, the probability that macroscopic current $ \cur^1_T$ takes value $ \cur^1$ can be proved to satisfy a large deviation principle \cite{qians}
\bea
P( \cur^1_T  \equiv  \cur^1) \asymp e^{-T I( \cur^1)}
\eea
where $\asymp$ means asymptotically in time and $I$ is the so-called rate function. Unfortunately, accessing the rate function of a special current $ \cur^1_T$ is a prohibitive task. Nevertheless, an exact result has been obtained by Maes et al. \cite{maesrate,wynants} (see also \cite{engel} for a pedagogical derivation) for the joint rate functional of $\jmath_T$ and $\rho_T$ (also see \cite{engel} for another perspective; for Markov jump-processes, finite-time corrections are available \cite{BEST}):
\begin{align}
I[\jmath,\rho] = \left\{\ba{ll}
\frac{1}{4} \int \rho^{-1} g^{-1}_{ij} (\jmath^i-\jmath^i_\rho) (\jmath^j-\jmath^j_\rho) &, \nabla j = 0 \\
+\infty &, \mathrm{otherwise}
\ea \right. .
\end{align}
It is crucial that the rate functional is only finite for a divergenceless current. The rate functional is non-negative. It only vanishes when $\jmath = \jmath_\rho$; taking the divergence, we obtain $\nabla \jmath_\rho =0$, which implies that $\rho = \rho_\star$ is the steady density and that $\jmath = \jmath_\star$ is the steady current. 

In principle, the rate function for the macroscopic current $ \cur_T^1$ can be obtained using the contraction principle,
\bea
I( \cur^1) = \inf_{\jmath | \langle \phi^1,\jmath \rangle =  \cur^1} I[\jmath], \label{eq:contr}
\eea
where $I[\jmath]$ is the rate functional for the currents, found by contracting over the density:
\bea
I[\jmath] = \inf_{\rho} I[\jmath,\rho] \label{eq:ik}.
\eea
We will be interested in the variance of $ \cur^1$, given by
\bea
\var  \cur^1 = \frac{1}{I''( \cur^1_\star)}
\eea 
where the prime denotes derivative with respect to $ \cur^1$. This identity is a consequence of the fact that $I( \cur^1)$ is the Legendre transform of the cumulant generating function of $ \cur^1_T$, and that the Legendre transform inverts the curvature \cite[p.\,20]{touchette}.

\subsection{Inequalities for the rate function: loose bound}

Eq.\,(\ref{eq:ik}) immediately implies the inequality
\bea
I[\jmath] \leq I[\jmath,\rho_\star] =: I_\star[\jmath], \qquad \nabla \jmath = 0, \label{eq:ineq}
\eea
that holds for any particular evaluation of $\rho$, in particular at the steady-state density $\rho_\star$. The right-hand side of the above equation defines the quadratic functional of the currents $I_\star$, which explicitly reads
\bea
I_\star[\jmath] = \frac{1}{4} \int \rho_\star^{-1} g^{-1}_{ij} (\jmath^i-\jmath^i_\star)  (\jmath^j-\jmath^j_\star).
\eea
Inequality (\ref{eq:ineq}) of course does not hold when $\jmath$ is not divergenceless, in which case $I[\jmath] =+\infty$. It is interesting to notice that this quadratic bound, found by Gingrich and coworkers \cite{gingrich}, is not the Gaussian approximation of the rate function around the steady state, as the second derivatives do not agree. Instead, as noted in  \cite{gingrich}, $I_\star$ is the parabola with the correct concavity which respects the Galavotti-Cohen symmetry. This already implies that the bound is saturated near equilibrium, where the two parabolas approach each other.   {See Sec.\,\ref{quadratic} for further insights about quadratic approximations of the rate function.}


Let us now consider an arbitrary macroscopic current $ \cur^1$, non-orthogonal. By Eq.\,(\ref{eq:contr}), $I( \cur^1_\star)$ is less than any evaluation of the rate functional of the microscopic currents that satisfies the constraints. A first bound is found by choosing
\bea
\jmath'_1 = \frac{ \cur^1}{ \cur^1_\star} \, \jmath_\star.
\eea
By construction, this choice automatically satisfies the linear constraints $\nabla \jmath'_ \cur =0$ and $\langle \phi^1,\jmath'_ \cur  \rangle =  \cur$. We then obtain the {\it loose bound} on the rate function
\bea
I( \cur^1)\leq I_\star(\jmath'_1) \leq \frac{\sigma_\star}{4} \left(\frac{ \cur^1}{ \cur^1_\star} -1 \right)^2. \label{eq:ine}
\eea

\subsection{Inequalities for the rate function: tight bound}

The above result holds for an arbitrary current. In this section we are going to show that there is a subclass of currents for which a tighter bound holds, and that this tighter bound can be interpreted in the light of the mininum entropy production principle.

To obtain a better bound, instead of considering fluctuations of the current that are proportional to the average current, a second choice is to pick the current that minimizes $I_\star(\jmath)$ at fixed $ \cur^1$, as was proposed in Ref.\,\cite{gingrich}:
\bea
\jmath_1 = \mathrm{arginf}_{\jmath | \langle \jmath,\phi^1\rangle =  \cur^1 } \;I_\star[\jmath]. 
\eea
This problem can be solved by simple linear algebra. We introduce a Lagrange multiplier $\lambda$ to keep into account the constraint, and impose that the constrained functional derivative of $I_\star[\jmath]$ with respect to $\jmath^i(x)$ vanishes: 
\bea
\frac{\delta}{\delta \jmath^i(x)} \left\{I_\star[\jmath] + \lambda \left(  \cur^1 - \langle \jmath, \phi^1 \rangle\right) \right\} = 0,
\eea
yielding
\bea
\jmath^i_1-\jmath^i_{\star} =  2\lambda \, \rho_\star g^{ij} \phi^1_j. \label{eq:sollam}
\eea
We remind that the above inequalities only hold for divergenceless currents, hence we need to impose that
\bea
0 = \nabla \jmath_1 = 2 \lambda \, \partial_i \left( \rho_\star g^{ij} \phi^1_j  \right). \label{eq:ssc}
\eea
This condition poses a constraint on $\phi^1$, thus restricting the set of macroscopic currents that obey the tighter bound that we are going to prove. We notice that this equation resembles Eq.\,(\ref{eq:sss}). It states that $\rho_\star$ is the steady state of both the complete system and of the system where only force $F^1 \neq 0$. In Appendix \ref{OU} we provide an example.

Next, we plug expression (\ref{eq:sollam}) into the constraint equation to solve for the Lagrange multiplier:
\bea
\lambda = \frac{ \cur^1- \cur^1_\star}{2\int \rho_\star g^{ij}  \phi^1_i \phi^1_j} 
\eea
so that
\bea
\jmath_1^i - \jmath^i_\star = \frac{ \cur^1- \cur^1_\star}{\int \rho_\star g^{i'j'} \phi^1_{i'} \phi^1_{j'}} \; \rho_\star g^{ij}  \phi^1_j.
\eea
Finally we derive the second main result of our paper, namely the {\it strict bound} for the rate function of a thermodynamically consistent current of the kind described above:
\begin{align}
I( \cur^1) \leq I_\star[\jmath_1] & = \frac{( \cur^1- \cur^1_\star)^2}{4 G^{11}} \nonumber \\ & = \frac{\sigma_\star^{11}}{4} \left(\frac{ \cur^1}{ \cur^1_\star} -1 \right)^2 \label{eq:tight}
\end{align}
where we recognized the nonequilibrium response coefficient $G^{11}$, and we introduced the {\it partial entropy production rate}
\begin{align}
\sigma_\star^{11} & := \frac{ ( \cur^1_\star)^2}{G^{11}}. \label{eq:tightert}
\end{align}
Now, given the definition of the quadratic dissipation function Eq.\,(\ref{eq:dissifun}), by simple linear algebra one can show that the partial entropy production rate is the infimum of the dissipation function, for fixed value of $\cur^1$:
\begin{align}
\sigma_\star^{11} = \inf_{\{ \cur^a\}_a |  \cur^1}  \sigma[\cur^1_\star,\ldots,\cur^n_\star].
\end{align}
In view of the discussion at the end of \S\,\ref{macros}, this quantity can be interpreted as the minimum entropy production rate that is compatible with an observed value of $ \cur^1$, for a perturbation of the steady currents that preserves the nonequilibrium response matrix, which implies that $\rho^\star$ remains unchanged. Therefore, the partial entropy production rate is that produced by a system that has the minimum possible entropy production rate that sustains current $ \cur^1$ on average, for a fixed steady density. This is the second key result of our paper.

A few comments are here in order. A bound analogous to that expressed in Eq.\,(\ref{eq:tightert}) has been provided in Ref.\,\cite[Eq.\,(16)]{gingrich} in the context of Markov jump processes, devoid of physical interpretation; there the matrix entry entering the bound is implicitly defined via pseudoinverse, which makes it dificult to compare the two results. It might be speculated that the tighter bound follows from the loose one, given that since it needs to hold for any system, it also has to hold for that system that has minimum entropy production rate. However, notice that Eq.\,(\ref{eq:tight}) compares the rate function of a certain system, which depends e.g. on the drift $\mu$, to the entropy production rate {\it of another system}, that has minimum entropy production. In fact, it is simple to verify that this new system has drift $\mu^i_1 = \mu^i - \sum_{a \neq 1} F_a g^{ij} \phi^a_j$. Therefore Eq.\,(\ref{eq:tight}) is not a trivial consequence of Eq.\,(\ref{eq:ine}). Furthermore, since the minimum entropy production principle lends itself to an information-theoretic understanding in terms of information that an observer has at his disposal about the system \cite{minepinfo}, then, in a way, $\sigma_\star^{11}$ is a good candidate as the measure of entropy production that an observer who only measures current $1$ could estimate. Notice that Eq.\,(\ref{eq:tight}) tightens the bound described in Refs.\,\cite{barato1,barato3,gingrich} in a way that still bears physical interpretation. 

It is an interesting exercise to re-derive the loose bound from the tight one. All is in place to employ the very same technique envisaged by Barato and Seifert to prove an analogous bound in the linear regime \cite[Supplementary Material]{barato1}. Let us expand all quantities in terms of the macroscopic forces:
\bes
( \cur^1_\star)^2 & = & G^{1a} G^{1b} F_a F_b  \\ 
G^{11} \sigma_\star & = & G^{11} G^{ab} F_a F_b .
\ees
Linear algebra tells us that matrix $(G^{11} G_{ab} -G^{1a} G^{1b})_{a,b}$ is positive definite, hence $ G^{11} \sigma_\star \geq ( \cur^1_\star)^2$
and Eq.(\ref{eq:ine}) follows. Finally, notice that if current $ \cur^1$ was orthogonal, we would have $ \cur^1_\star = 0$ but finite entropy production rate and finite variance, hence the error shoots to infinity and the bound would be trivially satisfied.

\subsection{Uncertainty relations} 

In Eqs.\,(\ref{eq:tight},\ref{eq:ine}), taking twice the derivative with respect to $ \cur^1$ and evaluating at $ \cur^1_\star$, given that $I( \cur^1_\star) = I_\star( \cur^1_\star) = 0$ and $I'( \cur^1_\star) = I'_\star( \cur^1_\star) = 0$, we obtain the hierarchy of inequalities
\bea
\frac{\var  \cur^1}{( \cur^1_\star)^2} \geq \frac{2}{ \sigma_\star^{11} } \geq \frac{2}{ \sigma_\star } \label{eq:tight2}.
\eea
One particular macroscopic current of interest is the entropy production rate $\sigma_T$ itself which, in view of Eq.\,(\ref{eq:EPR}), is selected by choosing $\phi^1 = f$ \cite{gingrich}. In this case a neat expression for the Fano factor of the entropy production rate is found, that we can write in compact form
\bea
\frac{\mathrm{var} \,\sigma}{\sigma_\star} \geq 2.
\eea

\section{Cycle currents of jump processes \label{schnak}}


To connect to Refs.\cite{barato1,barato3,gingrich}, and for sake of completeness, in this section we consider ergodic, continuous-time, discrete-state-space Markov jump processes, which occur on a network of states (a graph). It will soon be clear that the analysis above carries through in an analogous way, hence we do not repeat it to avoid redundancy. Nevertheless, we deem it interesting to inspect the theory in the light of Schnakenberg's analysis of cycle currents, which allows to automatically keep into account the steady-state constraint and provides a smooth transition from the formalism of nonequilibrium response functions to the linear regime.   {Note that Schnakenberg's decomposition of cycle currents plays an instrumental part in proving the non-quadratic bound discussed in Ref.\,\cite{pietzonka2}.}

\subsection{Setup}

Letting $e := x \gets y$ denote an oriented edge in the graph of the system, with $-e := x \to y$ the inverse edge, we introduce the steady semicurrents $k_\star^{+e} := w_{xy} \rho^\star_y$ and $k_\star^{-e} := w_{yx} \rho^\star_x$, where $w_{xy}$ is the transition rate and $\rho^\star_x$ the invariant measure.  The steady currents and their conjugate forces are defined as
\bes
\jmath^e_\star & := & k^{+e}_\star - k^{-e}_\star \\
f_e & := & \ln \frac{k^{+e}_\star}{k^{-e}_\star}.
\ees
Steady currents are divergenceless, that is, they satisfy $\nabla \jmath_\star = 0$, where $\nabla$ is the {\it incidence matrix} of the graph. The steady entropy production rate is the bilinear form \cite{schnak}
\bea
\sigma_\star = \sum_e \jmath^e_\star f_e = \langle\jmath_\star , f\rangle .
\eea

We now consider a stochastic realization of the currents $\jmath_T$ and in particular the rate function $I(\jmath)$. The following inequality has been proven in Ref.\,\cite{gingrich}:
\bea
I(\jmath) \leq I_\star(\jmath) := \frac{1}{4} \sum_e \left(\jmath^e - \jmath^e_\star\right)^2 \frac{f_e}{\jmath^e_\star}. \label{eq:ineqmjp}
\eea
Let us point out that the inequality   {only} holds on the assumption $\nabla \jmath = 0$. 

Finally, we consider one particular macroscopic current
\bea
\cur = \langle  \jmath,\phi \rangle
\eea
on the assumption that $\phi$ is not orthogonal to the steady current, so that $\cur_\star \neq 0$. It will be clear that, from now on, the treatment of the bounds on the rate functions and on the variances follows in the exact same way as in the previous section. A different perspective, though, is gained through the analysis of cycle currents, rather than of microscopic or of thermodynamically consistent currents.

\subsection{Cycle analysis}

Refs.\,\cite{barato1,barato3} mainly refer to Schnakenberg's cycle currents, which are solutions to the divergence equation $\nabla \jmath =0$. The analysis of large deviations proposed in \cite{wachtel} states that only cyclic terms contribute to the full statistics of the currents. In Schnakenberg's formalism there naturally emerges a nonequilibrium response matrix for the cycle currents, which allows to prove that the bound for the entropy production rate saturates in the linear regime in a straightforward manner.

Equation $\nabla \jmath =0$ implies that currents live in the kernel of the incidence matrix, which is spanned by independent {\it cycle vectors} $(c^\alpha_e)_e$. Schnakenberg's theory basically consists in enforcing this condition (and in choosing a preferred basis of cycles generated by a spanning tree, whose structure is here irrelevant):
\bea
\jmath_\star  = c_a\, \mathcal{J}_\star^a. \label{eq:cyclecurr}
\eea
Let us define the (inverse) {\it nonequilibrium response function}
\bea
\mathcal{G}^{-1}_{ab} := \sum_e c^e_a c_b^e \frac{f_e}{\jmath_\star^e},
\eea
such that
\bea
 \mathcal{J}_\star^a = \mathcal{G}^{ab} \mathcal{F}_b
\eea
where the cycle forces are defined as
\bea
\mathcal{F}_a := \langle c_a , f\rangle.
\eea
The entropy production can then be expressed in terms of cycle observables as
\bea
\sigma_\star = \mathcal{J}^a_\star \mathcal{F}_a =  \mathcal{G}^{ab} \mathcal{F}_a \mathcal{F}_b .
\eea
Close to equilibium, matrix $\mathcal{G}_{ab}$ coincides with the linear response matrix described in \cite{schnak}, which finds application for example in the proper formulation of the minimum entropy production principle \cite{polettiniminEP}.

We can now express the physical current as
\bea
\cur = \Phi_a \mathcal{J}^a,
\eea
where $\Phi_a  =  \phi \centerdot  c_a$.

Employing the fact that not all currents are independent, we can contract the latter inequality to the cycle currents by simply replacing $\jmath = c_a \mathcal{J}^a$ in Eq.\,(\ref{eq:ineqmjp}),
\bea
I( \mathcal{J}) \leq I_\star( \mathcal{J})
\eea
where
\bea
I_\star( \mathcal{J}) = \frac{1}{4} ( \mathcal{G}^{-1}_{ab} \mathcal{J}^a \mathcal{J}^b - 2 \mathcal{J}^a \mathcal{F}_a + \sigma_\star ). 
\eea
We then have
\bea I(\cur) \leq I_\star( \mathcal{J}_\cur) \eea
where $ \mathcal{J}_\cur$ is the infimum  of $I_\star( \mathcal{J})$ for a fixed value of $\cur$, which is given by \footnote{We notice that, if we minimized $I(\jmath)$ at fixed $\cur$ with respect to $\jmath$, the solution $\jmath^\cur$ would not generally be divergenceless, hence it would fall out of the domain of applicability of inequality (\ref{eq:ineqmjp}).}
\bea
 \mathcal{J}_{\cur}^a  = \mathcal{G}^{ab} \left( \frac{\cur - \cur_\star}{\mathcal{G}^{a'b'} \Phi_{a'} \Phi_{b'}}  \Phi_b  +  \mathcal{F}_b\right).
\eea
We now evaluate
\bea
I_\star( \mathcal{J}_\cur) & = &  \frac{(\cur - \cur_\star)^2}{4\mathcal{G}^{ab} \Phi_a \Phi_b},
\eea
which we believe to be a more explicit version of Eq.\,(16) from \cite{gingrich}, and in particular taking the second derivative and evaluating at $\cur_\star$ we obtain the tight bound
\bea
\frac{\var \cur}{\mathcal{G}^{ab} \Phi_a \Phi_b}  \geq 2. 
\eea
In particular when $\Phi_a = \mathcal{F}_a$, we obtain the entropy production bound Eq.\,(\ref{eq:eprbound}). However, differing from the case of the thermodynamically-consistent current that lead to the tighter bound Eq.\,(\ref{eq:tight}), in this case there is no immediate physical interpretation for $\mathcal{G}^{ab} \Phi_a \Phi_b$ in terms of a partial entropy production rate. At this point, introducing a thermodynamic consistency condition would lead us to the tight bound discussed above. We will not repeat the discussion.

The loose bound Eq.\,(\ref{eq:eprbound}) can be obtained as follows. Since $\mathcal{G}^{ab}$ is symmetric positive-definite, it is Gramian: there exists a ``square root'' matrix  $\ell^a_i$ (with inverse $\ell_a^i$) such that
\bea
\mathcal{G}^{ab} =  \langle \ell^a,\ell^b \rangle.
\eea
where $\langle\cdot,\cdot\rangle$ denotes the Euclidean scalar product. Then
\bea
\mathcal{G}^{ab} \Phi_a \Phi_b & = & \langle\ell^a \Phi_a, \ell^b \Phi_b\rangle \\
\sigma_\star & = & \langle\ell_a  \mathcal{J}_\star^a, \ell_b  \mathcal{J}_\star^b\rangle
\eea
and by the Cauchy-Schwarz inequality
\bea
\cur_\star^2 = \langle \ell^a \Phi_a,\ell_b J_\star^b \rangle^2 \leq \sigma_\star \mathcal{G}^{ab} \Phi_a \Phi_b   \label{eq:bb}
\eea
Then:
\bea
\frac{\var \cur}{\cur_\star^2} \sigma_\star \geq \frac{\var \cur}{\mathcal{G}^{ab} \Phi_a \Phi_b}  \geq 2.
\eea
  {This is the analogue of Eq.\,(\ref{eq:tight2}) for Markov jump processes. The interpretation in terms of the minimum rate of entropy produced by a system that has the same response matrix can also be retraced. However, we notice in passing that, while for diffusion processes the response matrix is determined in terms of the diffusion tensor and the steady-state distribution, in this case the response matrix is a rather {\it ad-hoc} object involving a very special combination of steady-state currents and forces; it is not obvious a priori what kind of transformations of the transition rates of the system will preserve the response matrix.}

\subsection{Optimality of the bound}

In the linear regime the bound for the entropy production rate saturates. In fact using the Green-Kubo relations we obtain 
\begin{align}
\var \sigma = \mathcal{F}_a \mathcal{F}_b \mathrm{cov} (\mathcal{J}^a,\mathcal{J}^b) = 2 \mathcal{F}_a \mathcal{F}_b \mathcal{G}^{ab} = 2\sigma_\star .
\end{align}
Then, at least close to equilibrium, the entropy production rate is ``optimal'', in the sense that any other current performs worse. Let us then inquire whether the entropy production rate is always the physical current that optimizes the bound. We investigate this question with a simple model study, finding that as one goes far fom equilibrium, deviations from optimality of the entropy production rate are small.  

We consider a Markov jump process on the four-state network
with rates $w_{+1} = w_{+2} = w_{+3}= w$, $w_{-1}= w_{-2}= w_{-3}= w_{+5} = w_{-5} = 1$, $w_{+4} = 2w$, $w_{-4} =2$ in terms of the driving parameter $w$. The affinities are given by
\bea
A_1 = A_2 = A = 2\log w
\eea
and the system appoaches equilibrium for $w \to 1$. We consider a current in the form
\bea
\cur = A \mathcal{J}^1 + x \mathcal{J}^2
\eea
which for $x = A$ corresponds to the entropy production rate. We calculate
$f(x) = \sigma_\star \var \phi / \cur_\star^2$ as a function of $x$. For $w=2$, Fig.\,\ref{4}, shows that $f(x)$ approaches the optimal bound for some value of $x$. We calculate $x^{\mathrm{opt}}$ for which the bound is optimized, and confront it to the affinity $A$. We find that these values are very close and that they get closer as $w \to 1$, as shown in the plot in Fig.\,\ref{1}. Furthemore, Fig.\,\ref{2} shows that the optimal error, relative to the theoretical value $2$, is approached as $w\to 1$, and that there is almost no difference in error between the optimal current and the entropy production rate. In the range of $w$ we considered, the entropy production rate spans two orders of magnitude.
\begin{figure}
  \centering
   \includegraphics[width=\columnwidth]{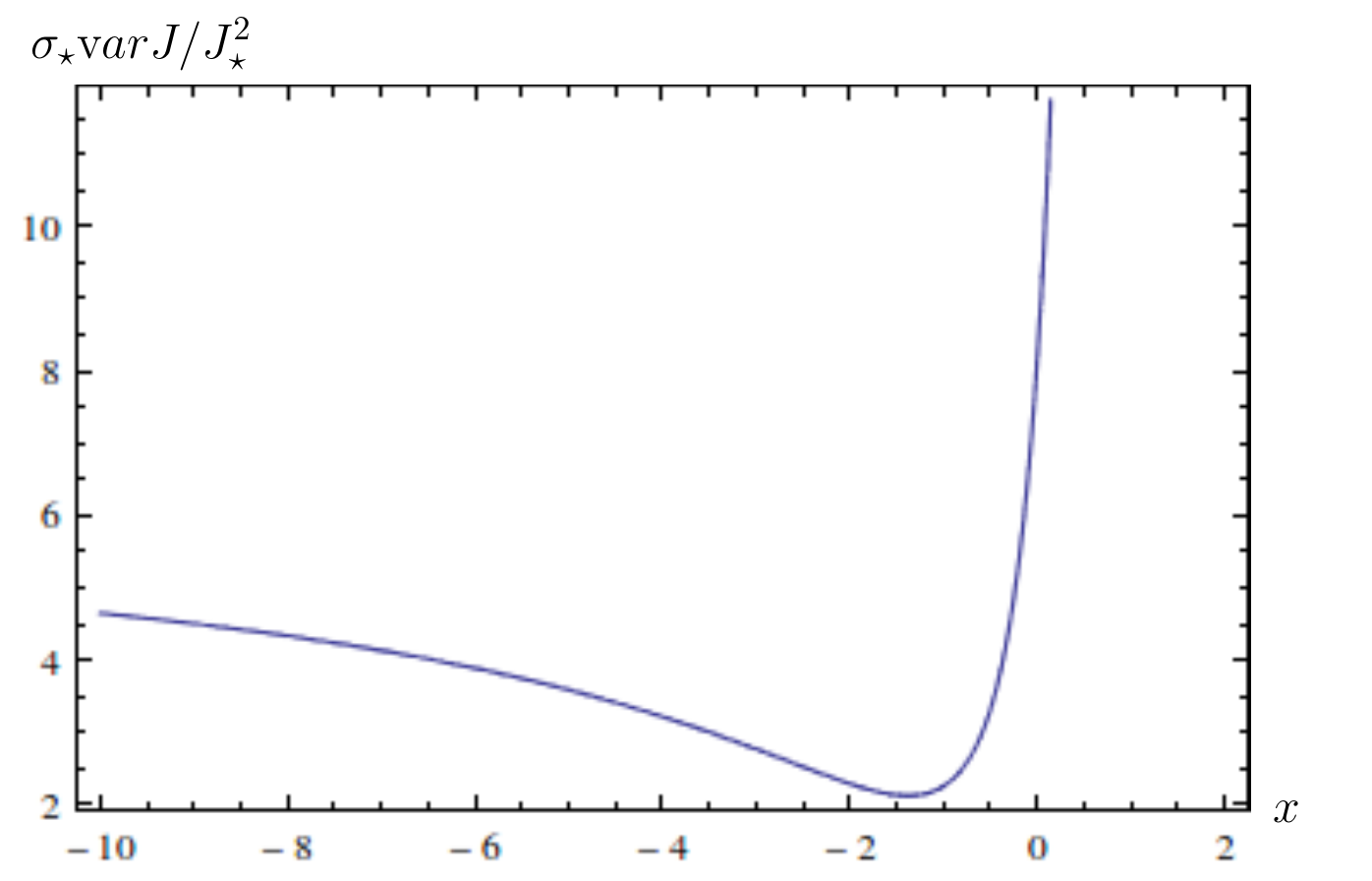}
  \caption{\label{4} The squared error of the current $f(x) = \sigma_\star \var \cur / \cur_\star^2$, as a function of parameter $x$. The plot shows that there is a value of $x^{\mathrm{opt}}$ for which the bound is optimal.}
\end{figure}

\begin{figure}
  \centering
   \includegraphics[width=\columnwidth]{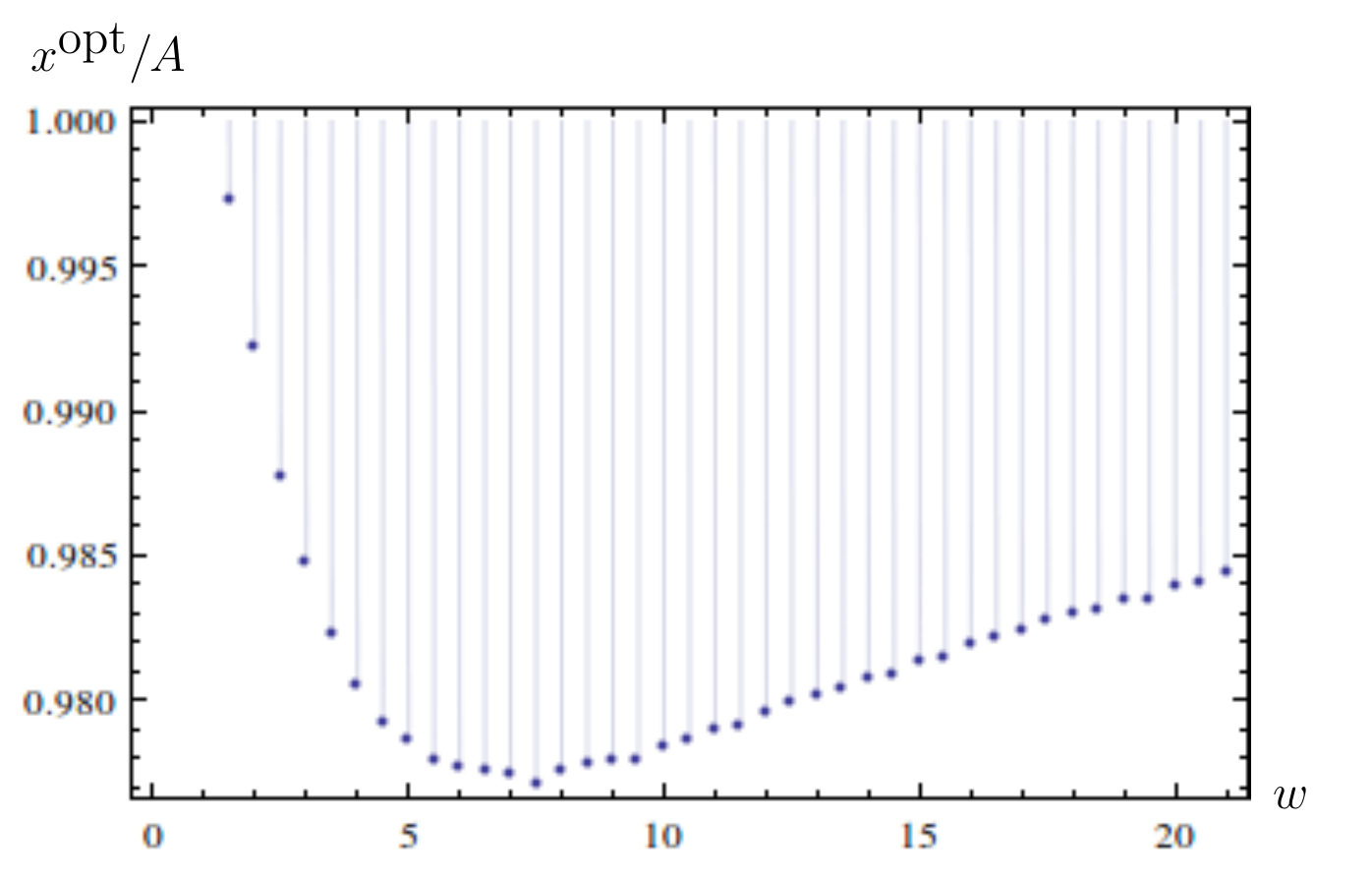}
  \caption{\label{1} The relative optimal affinity $x^{\mathrm{opt}}/A$ as a function of the driving parameter $w$.}
\end{figure}

\begin{figure}
  \centering
   \includegraphics[width=\columnwidth]{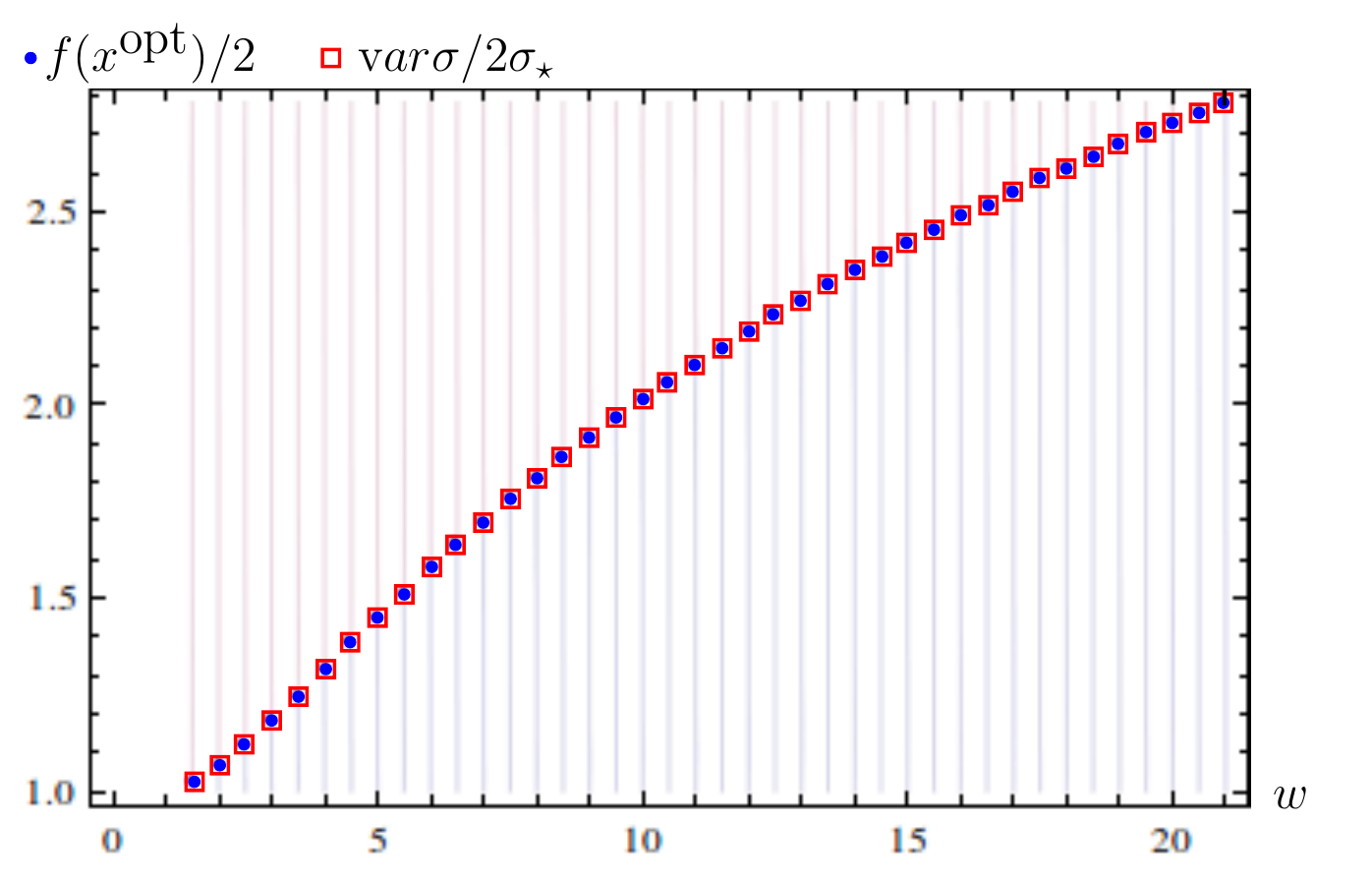}
  \caption{\label{2} In this plot two set of overlapping points are plotted. The set of circles corresponds to the relative optimal error  $f(x^\mathrm{opt})/2$; the set of squares correspond to the relative entropy production rate Fano factor $\var \sigma/(2\sigma_\star)$, both plotted as a function of the driving parameter $w$. The image clearly shows that the bound tends to perform worse far from equilibrium, and that there is almost no difference between the optimal current and the entropy production rate.}
\end{figure}

Another example of the non-optimality of that bound can be seen at the end of the Appendix for a system with three edges but only two independent cycles.


\section{Conclusions}

In this paper we discussed the uncertainty relation for the currents recently discovered by Barato and Seifert \cite{barato1,barato2,barato3} and proved by Gingrich et al. \cite{gingrich}. We first examined the conditions for the appearance of such relations in stochastic processes, including Markov jump processes and diffusions, and provided a simple proof of the inequality. We then focus on overdamped diffusion processes, finding that a notion of {\it thermodynamic consistency} and of {\it symmetry} of the thermodynamic forces is useful to produce and interpret a tighter bound on a class of physical currents, in terms of the least possible entropy production rate that is compatible with the observed value of the current, and with the steady density. A notion of {\it nonequilibrium response function} naturally emerges from our treatment. We then performed a similar analysis in the case of Markov jump processes, employing Schnakenberg's theory of cycle currents, which allows to clarify in which sense is the entropy production rate the optimal current with respect to the loose bound. In the future it might be interesting to connect this theory to other results concerning the Fano factor of the heat in interacting particle models \cite{derrida}.

\section*{Acknowledgments}

The authors are grateful to Jordan Horowitz and Hugo Touchette for helpful discussions, comments, and for pointing out some errors in a previous version of the manuscript. The research was supported by the National Research Fund Luxembourg in the frame of project FNR/A11/02, and by the European Research Council (project 681456). AL was supported by the AFR PDR 2014-2 Grant n$^{\circ}$.9202381.

\appendix

\appendix

\section{Optimal quadratic bounds on the currents}

\label{appendix2}

In order to optimise the quadratic bound on $F(j)$, we start by reducing the problem by one dimension.

Asking that $F(\jmath)-\frac{1}{4}\sigma^\star\leq\langle\jmath, A \jmath\rangle$ is equivalent to asking that the level manifolds of the right-hand side lie inside of those of the left-hand side. That is to say that, for any $a\in \mathbb{R}$, with $\frac{1}{4}\sigma^\star$ chosen as a natural scale for $F$:
\begin{equation}
\biggl\{\jmath\bigg|F(\jmath)-\frac{1}{4}\sigma^\star\leq a\frac{1}{4}\sigma^\star\biggr\}\supseteq\biggl\{\jmath\bigg|\langle\jmath, A \jmath\rangle\leq a\frac{1}{4}\sigma^\star\biggr\}.
\end{equation}
Note that each of the sets in the left-hand side are convex, because $F$ is convex. We can simplify greatly this expression by noticing that the right-hand side always gives the same set up to a rescaling by $\sqrt{a}$. We can then rewrite the conditions so as to have the same right-hand side, and regroup the left-hand sides into
\bea\label{condS}
S & = & \bigcap\limits_{a\in \mathbb{R}}\biggl\{\jmath\bigg|F(\jmath\sqrt{a})-\frac{1}{4}\sigma^\star\leq a\frac{1}{4}\sigma^\star\biggr\} \nonumber \\
&\supseteq& \biggl\{\jmath\bigg|\langle\jmath, A \jmath\rangle\leq \frac{1}{4}\sigma^\star\biggr\}.
\eea
The problem of finding an appropriate quadratic bound then reduces to finding a metric $A$ such that the ball of radius $\frac{1}{4}\sigma^\star$ is contained in $S$. This set is an intersection of convex sets, so it is convex itself. Moreover, it has the Gallavotti-Cohen symmetry: if $\jmath$ is in $S$, then so is $-\jmath$.

Note that, as required, $\jmath^\star$ is on the boundary of $S$: this is ensured by the constant $\frac{1}{4}\sigma^\star$ removed from $F(j)$, setting a reference for the level sets at its value in the stationary state, and by the fact that $\alpha\frac{{\rm d}^3}{{\rm d}\alpha^3}F(\alpha \jmath^\star)\leq 0$.

~~

Obtaining an optimal solution is then entirely problem-dependent, and there is typically a continuous set of candidates. Luckily, there is a constructive way to obtain them. A current $\jmath$ can be decomposed onto $j^\star$ and the space orthogonal to it with respect to the metric $A$: $\jmath=\alpha \jmath^\star+\jmath'$ with $\langle\jmath^\star, A \jmath'\rangle=0$. We then have $\langle\jmath, A \jmath\rangle=\alpha^2\frac{\sigma^\star}{4}+\langle\jmath', A \jmath'\rangle$, and the condition given in Eq.\,(\ref{condS}) becomes
\begin{multline}
\bigcap\limits_{a\in \mathbb{R}}\biggl\{\jmath'\bigg|F((\alpha \jmath^\star+\jmath')\sqrt{a})-\frac{1}{4}\sigma^\star\leq a\frac{1}{4}\sigma^\star\biggr\}  \\ \supseteq\biggl\{\jmath'\bigg|\langle\jmath', A \jmath'\rangle\leq \frac{1-\alpha^2}{4}\sigma^\star\biggr\}.
\end{multline}
We can, once more, rescale the right-hand side and regroup the left-hand sides, to get a new constraint on a smaller space:
\begin{align}\label{condS'}
& S' = \nonumber \\
& \bigcap\limits_{\substack{a\in \mathbb{R} \\ \alpha\in[0,1]}}\biggl\{\jmath\bigg|F((\alpha \jmath^\star+\sqrt{1-\alpha^2}\jmath)\sqrt{a})-\frac{1}{4}\sigma^\star\leq a\frac{1}{4}\sigma^\star\biggr\}  \nonumber \\
& \supseteq \biggl\{\jmath\bigg|\langle\jmath, A \jmath\rangle\leq \frac{1}{4}\sigma^\star\biggr\}
\end{align}
where $A$ and $\jmath$ are now restricted to the space orthogonal to $f^\star$.

\begin{figure}[b]
  \centering
   \includegraphics[width=\columnwidth]{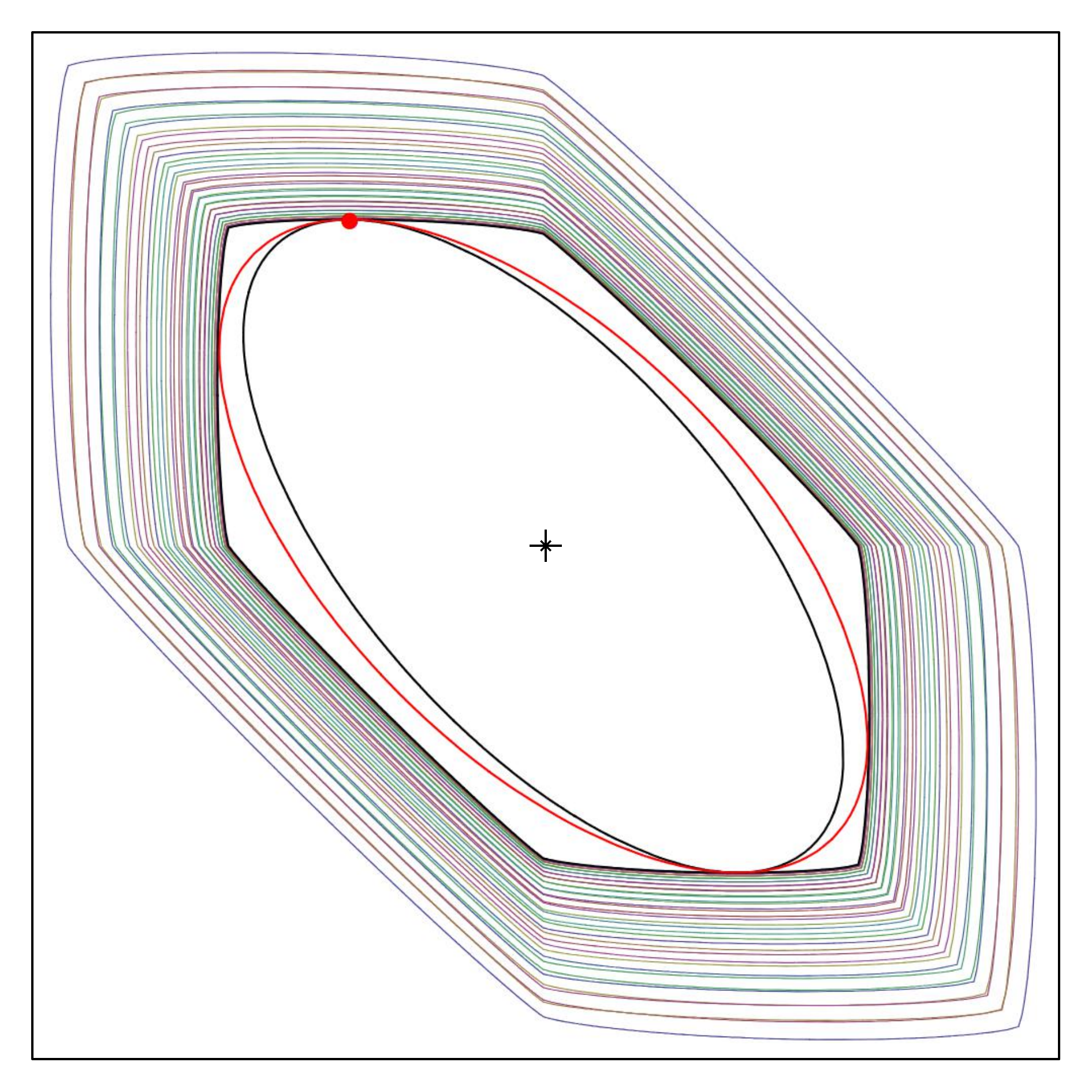}
  \caption{Rescaled and centred level curves of the large deviation function of the stationary currents in a two-state/three-channels model. The stationary currents are marked by the red dot. The outer (red) ellipsis gives the optimal quadratic bound on that function, and the inner (black) ellipsis gives the one which is diagonal in the basis of edges.}
\end{figure}

This process can be repeated until $A$ is completely determined. However, since we now have no \textit{a priori} preferred choice for a point where the inclusion should saturate, we have to choose a point on the boundary of $S'$ at every step, which produces a continuous set of solutions. Moreover, every step gives us an extra saturation point for the inclusion constraint, unless the set for $\alpha=\pm 1$ is the smallest one, in which case we have that the curvature of the two sides becomes the same at the corresponding point. At the end of the procedure, we therefore have a number of constraints, be it saturation or equal curvature at saturation, equal to 2$d$, where $d$ is the dimension of the cycle current space (the factor $2$ comes from the Gallavotti-Cohen symmetry).

As an illustration, let us look at a very simple model with two states connected by three channels. This $\jmath_1$, $\jmath_2$ and $\jmath_3$, with the stationarity condition $\jmath_1+\jmath_2+\jmath_3=0$. In the following figure, we plot, as functions of $\jmath_1$ and $\jmath_2$, the solutions of $F(\jmath\sqrt{a})-\frac{1}{4}\sigma^\star=a\frac{1}{4}\sigma^\star$ for various values of $a$ (coloured hexagonal lines), the ellipsis corresponding to the optimal quadratic bound (red), that for the edge-wise bound (black), and the value of the average currents (red dot). As can be seen, the optimal bound saturates at four values of the current, and is strictly more precise that the edge one.

\section{The Ornstein-Uhlenbeck process}
\label{OU}

Let us show by an example the nature of the class of currents for which the tighter bound holds. We consider an Ornstein-Uhlenbeck process
\bea
dx^i = - \Gamma^i_j x^j \, dt + dw^i_t,
\eea
with $g_{ij} = \delta_{ij}$ and $\Gamma$ a positive-definite matrix. The Fokker-Planck equation reads
\bea
\partial_t \rho_t = \nabla (\Gamma x \rho_t + \nabla \rho_t),
\eea
and the steady ensemble is given by
\bea
\rho_\star \propto \exp - \frac{1}{2} \Sigma^{-1}_{ij} x^i x^j,
\eea
where the covariance matrix is determined in terms of $\Gamma$ via the equation  \cite{risken}
\bea
\Sigma \Gamma^\dagger + \Gamma \Sigma = 2 \mathrm{I}, \label{eq:eva}
\eea
where $\mathrm{I}$ is the identity matrix. As generic macroscopic currents we consider
\bea
\cur^a = \int dx \, \Theta^a_{ij} x^i   \jmath^j(x), \label{eq:justi}
\eea
that is $\phi^a_j = x^i \Theta^a_{ij}$ for a collection of matrices $\Theta^a$.  Notice that the steady-state thermodynamic force reads $f = \jmath_\star / \rho_\star = (\Gamma - \Sigma^{-1}) x$. Thermodynamic consistency is granted provided that the collection of matrices $\Theta^a$ is complete in the sense that the linear system $\Gamma - \Sigma^{-1} = F^a {\Theta^a}^\dagger$ admits solutions; if it does not, then the set of macroscopic currents we are considering are not sufficient. Notice that the possibility of realizing thermodynamic consistency in this kind of system relies on the fact that the steady-state thermodynamic force is linear in $x$, which justifies the definition of the macroscopic currents Eq.\,(\ref{eq:justi}). Any functional that is not linear in $x$ will fail in this respect.

Let us now focus on the first such current $a=1$. We now need to impose Eq.\,(\ref{eq:ssc}), which yields the two conditions on $\Theta^1$
\bea
\mathrm{tr}\, \Theta^1 & = 0, \\
 \Sigma^{-1} {\Theta^1}^\dagger + \Theta^1  \Sigma^{-1} & = 0.
\eea

Let us look at some specific cases. First we consider
\bes
dx^1 & = & - (x^1 + x^2) dt + dw^1_t \\
dx^2 & = & - (x^2 - x^1)dt + dw^2_t
\ees
yielding
\bea
\Gamma = \left( \ba{cc} 1 & 1 \\ -1 & 1 \ea \right), \Sigma = \left(\ba{cc} 1 & 0 \\ 0 & 1 \ea \right), \Theta^1 = \left(\ba{cc} 0 & 1 \\ -1 & 0 \ea \right) 
\eea
and
\bea
\cur^1 = \int dx_1 \int dx_2 \, [ x^1   \jmath^2(x) -  x^2   \jmath^1(x)].
\eea
The thermodynamic force is
\bea
f(x) = \rho_\star^{-1}(x) \jmath_\star(x) = \left(\ba{c}- x_2 \\ x_1\ea \right)
\eea
hence in this case the only macroscopic current that satisfies the tighter uncertainty principle $J^1$ is actually the entropy production rate itself, and the tight bound reduces to the loose one. This is due to the fact that with two degees of freedom there is only one current.

Let us then move to three degrees of freedom. We choose
\bea
\Gamma = \left( \ba{ccc} 1 & 1 & 1 \\ -1 & 1 & 1 \\ -1 & -1  & 1  \ea \right)
\eea
yielding $\Sigma_{i,j} = \delta_{i,j}$. Again for $\Theta^1$ we can choose an arbitrary skew-symmetric matrix, for example
\bea
\Theta^1 = \left( \ba{ccc} 0 & 1 & 0 \\ -1 & 0 & 0 \\ 0 & 0 & 0  \ea \right).
\eea
In this case the macroscopic current reads
\bea
\cur^1 = \int dx_1 \int dx_2 \int dx_3 \, [ x^1   \jmath^2(x) -  x^2   \jmath^1(x)],
\eea
which is strictly different than the entropy production rate, since at the steady state we have that $2 = J^1_\star = \sigma_\star/3$. Indeed, $\cur^1$ in this case can be interpreted as an independent component of the total entropy production rate. Let us now determine the system that has minimum entropy production rate compatible with the steady state and the observed value of the current. Let its drift be $\mu^1(x)$. From the steady-state equation  $0 = \nabla (\Gamma x \rho_\star + \nabla \rho_\star)$ one immediately concludes that $\mu_1(x)$ must be linear $\mu_1(x) = - \Gamma_{1} x$, therefore we remain within the class of OU processes. From Eq.\,(\ref{eq:eva}) it follows that
\bea
\Gamma_{1} =  \left( \ba{ccc} 1 & a & b \\ -a & 1 & c \\ -b & -c  & 1  \ea \right)
\eea
and a straightfoward evaluation of the entropy production rate yields $\sigma_\star = 2 a^2 + 2b^2 + 2 c^2$ while 
$J^1 = 2a^2$, 
hence as could be expected it is straightfoward that the minEP system that sustains the current $J^1 = 2$ is the one with $a=1$, $b=c=0$. We can therefore conclude that the tight bound for the current's variance is three times stricter than the loose bound; however, we are not aware of simple techniques to actually perform a direct calculation for OU processes of the current's variance.

\end{document}